\documentclass[10pt,twocolumn,twoside]{IEEEtran}
\usepackage{amsmath,amsfonts}
\usepackage{algorithmic}
\usepackage{algorithm}
\usepackage{array}
\usepackage[caption=false,font=normalsize,labelfont=sf,textfont=sf]{subfig}
\usepackage{textcomp}
\usepackage{stfloats}
\usepackage{url}
\usepackage{verbatim}
\usepackage{graphicx}
\usepackage{cite}
\usepackage{booktabs}
\usepackage{enumitem}
\usepackage{bm}
\usepackage{multirow}
\usepackage{caption}
\usepackage{mathrsfs}
\usepackage{url}
 
\DeclareMathOperator{\multihead}    {Multihead}

\DeclareMathOperator{\head}{head}
\DeclareMathOperator{\softmax}{Softmax}

\DeclareMathOperator{\concat}{Concat}

\DeclareMathOperator{\onehot}{Onehot}

\begin{document}
\title{
Guided Masked Self-Distillation Modeling
\\for Distributed Multimedia Sensor Event Analysis

}
\author{Masahiro Yasuda,
Noboru Harada,\\
Yasunori Ohishi,
Shoichiro Saito,
Akira Nakayama,
Nobutaka Ono
\thanks{Masahiro Yasuda, Noboru Harada, Yasunori Ohishi, Akira Nakayama, and Shoichiro Saito are with NTT Corporation, Tokyo, Japan.}
\thanks{Masahiro Yasuda and Nobutaka Ono are with Tokyo Metropolitan University, Tokyo, Japan.}
}

\markboth{}{}

\maketitle
\begin{abstract}
Observations with distributed sensors are essential in analyzing a series of human and machine activities (referred to as 'events' in this paper) in complex and extensive real-world environments. This is because the information obtained from a single sensor is often missing or fragmented in such an environment; observations from multiple locations and modalities should be integrated to analyze events comprehensively. However, a learning method has yet to be established to extract joint representations that effectively combine such distributed observations. Therefore, we propose Guided Masked sELf-Distillation modeling (Guided-MELD) for inter-sensor relationship modeling. The basic idea of Guided-MELD is to learn to supplement the information from the masked sensor with information from other sensors needed to detect the event. Guided-MELD is expected to enable the system to effectively distill the fragmented or redundant target event information obtained by the sensors without being overly dependent on any specific sensors. To validate the effectiveness of the proposed method in novel tasks of distributed multimedia sensor event analysis, we recorded two new datasets that fit the problem setting: MM-Store~\footnote{\scriptsize{Sample data for peer review can be accessed here: \url{https://drive.google.com/drive/folders/19cxeWc18ian6TxL14VdwzDk97z9IjBsr?usp=sharing}. The full version will be available after acceptance.}} and MM-Office. These datasets consist of human activities in a convenience store and an office, recorded using distributed cameras and microphones. Experimental results on these datasets show that the proposed Guided-MELD improves event tagging and detection performance and outperforms conventional inter-sensor relationship modeling methods. Furthermore, the proposed method performed robustly even when sensors were reduced.
\end{abstract}

\begin{IEEEkeywords}
Event detection, Distributed camera, Distributed microphone, Cross-modal action recognition, Multi-view action recognition, Masked signal modeling
\end{IEEEkeywords}
\section{Introduction}
\label{sec:intro}
\begin{figure*}[t!]
 \centering
 \includegraphics[width=0.9\linewidth]{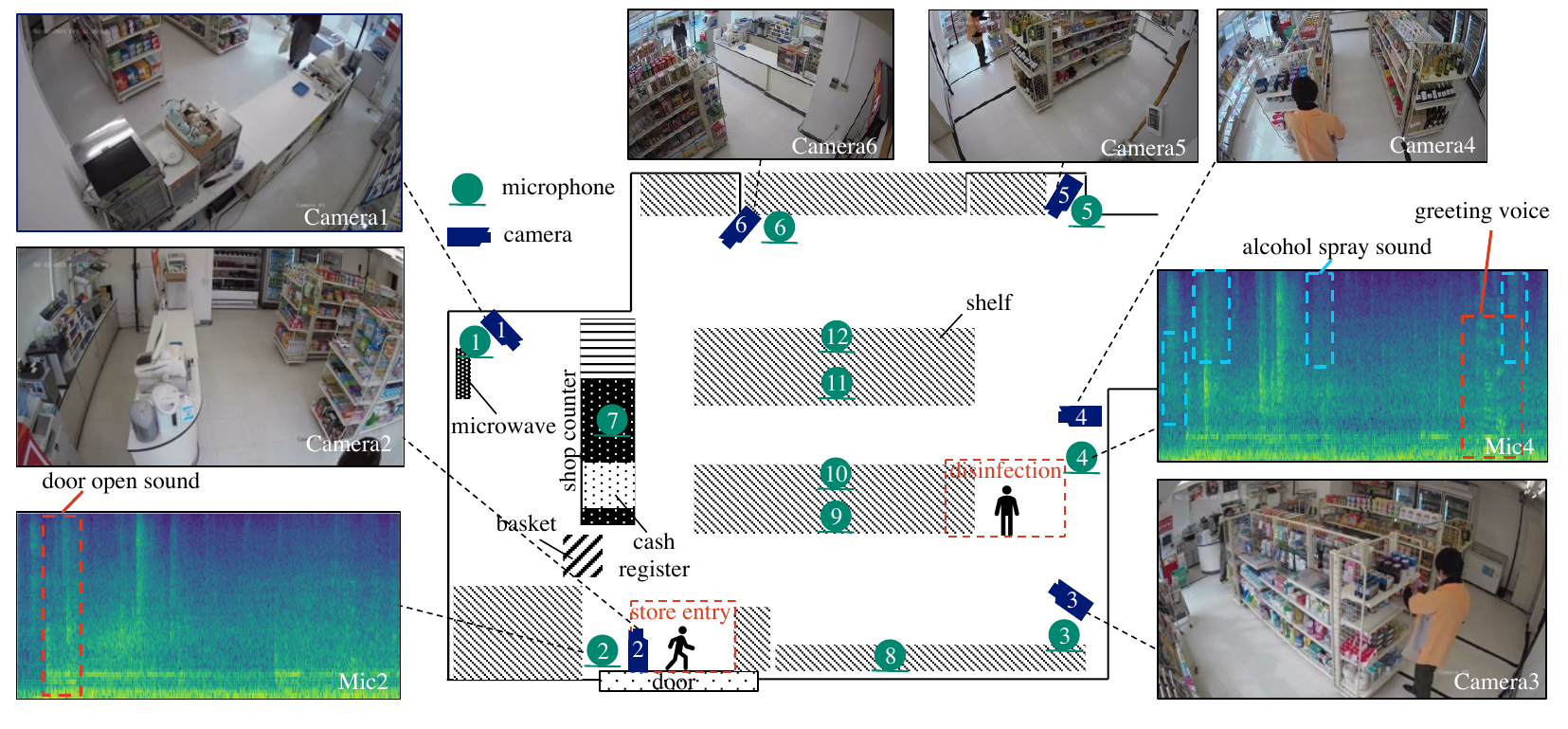}
\caption{An example of the observation for the distributed multimedia sensor event analysis task in our dataset: MM-Store. \\ 
{\footnotesize This example shows a scene in which a store staff member disinfects a store with an alcohol spray while a customer enters through the entrance. Camera 3 clearly captures the clerk, but his hand is difficult to see. On the other hand, microphone 4 clearly captures the sound of the alcohol spray being sprayed. The distributed multimedia sensor event analysis task requires distilling useful information for event identification from such observations. This figure also shows the room and sensor setup for the MM-Store dataset (details are given in Section V-A).}}
 \label{fig:mmstore}
\end{figure*}

This paper addresses a new task: distributed multimedia sensor event analysis (DiMSEA). The input signals in this task are multi-channel and multi-modal sensor signals. As Figure~\ref{fig:mmstore} shows, this study focuses particularly on the use of cameras and microphones distributed throughout the room as sensors. This task is broken down into two subtasks: event tagging and event detection. The former identifies the class of events contained in a recording, and the latter, in addition, identifies the onset/offset time of each event. Here, events are defined as a series of human actions or machine movements. For multiple events occurring simultaneously, tagging and detection is performed for each in parallel. 
As a practical constraint for annotation cost, this study approaches DiMSEA under the condition that only weak labels without event time annotations are available.

One promising application of this technology is automatically detecting customer behavior in stores. In the store environment, capturing information over the entire area with only a single sensor is difficult due to obstacles such as shelves and acoustic noise. Successful coordination of distributed multimedia sensors is expected to enable the analysis of target events even in such situations. By automatically detecting customer behavior, this system will make it possible to maintain and improve the quality of customer service and security, even if the store has few staff members or is unmanned. Other possible applications include wide-area security systems that detect suspicious persons and signs of violence~\cite{multi5} or crime~\cite{mvc}, house cleaning systems that automatically disinfect areas touched by people to prevent the spread of infectious diseases, and automatic work support systems for employees in offices. In anticipation of such applications, this study recorded datasets for convenience stores and offices. 

The primary research question in DiMSEA is how to extract information related to the target event from observations with distributed multimedia sensors. In DiMSEA, distributed sensors are deployed throughout the entire target space to ensure broad coverage. On the other hand, Target events, occur locally in this space. It implies that the observation signals from distributed sensors contain numerous background signals, and the identification of which sensors are observing the target event varies depending on the situation. In addition, the target event is observed fragmentarily by multiple sensors. These fragmented observations include clues about the event that are redundant across sensors, as well as clues unique to a sensor. Previous studies have pointed out that the validity of which viewpoints and modalities of cues depend on the class of events~\cite{mva4,multi1,multi5,multi6}. Therefore, for a system to effectively utilize the observations of distributed sensors in event analysis, the information of the target event observed fragmentarily among numerous background signals should be extracted by modeling inter-sensor relationships that enable the complementary combination of these observations.

Although the cooperation of multiple sensors has been utilized in previous studies for action recognition~\cite{mva4,mva5,mmact,avdataset1} and autonomous driving~\cite{crossview, deepfusion}, there are limitations in applying these methods to DiMSEA. One established method for coordinating multiple sensors is the conditional random fields (CRF)-based method~\cite{danet, crfpaper}, which models sensor relationships with a fixed weight matrix. However, such systems fail in model scenarios like DiMSEA, where the useful sensors change depending on the situation. Our previous study addressed this issue by utilizing MultiTrans~\cite{multitrans}, a Transformer-based system capable of paying attention to the necessary sensors on the basis of input~\cite{attention}. Nevertheless, even with this input-dependent attention mechanism, there remains a problem where the trained system erroneously focuses on a few specific sensors due to the training data is practically insufficient for the vast number of situations that can be observed by distributed sensors.

In the field of representation learning for tasks that analyze signals from single sensors such as images or sound, self-supervised learning methods have gained attention for modeling such signals. 
In particular, masked signal modeling (MSM) has been successfully utilized for representation learning of a wide range of modalities, including image, audio, acoustic, and language~\cite{cmmasksd,dm2,mam,mim2,msm1,mlm1,mlm2,m2d}.
Although no prior work that utilizes MSM to model inter-sensor relationships found, its ability to model the properties of signals well, regardless of modality, would be promising for DiMSEA.

However, it is reported that MSM-based representation learning is not robust for background signal~\cite{ask2mask, attnmsm}. In other words, MSM does not distinguish whether the input signal is an object of interest. This is a fatal drawback in modeling the distributed sensor signals that contain many background.
Indeed, in our preliminary investigations, a naive adaptation of MSM fails to learn a joint representation of distributed multimedia sensors (this will be discussed in Sec.~\ref{sec:resultstore}).

We propose Guided Masked sELf-Distillation modeling (Guided-MELD) to make distributed sensors cooperate effectively. 
Guided-MELD extracts a joint representation that distills information relevant to target events while excluding disturbing background. The key of Guided-MELD are twofold. First, it ensures that embeddings extracted from any subset of distributed sensor systems are as consistent as possible. It enables the system to extract the maximum possible information regardless of the specific sensors used. The second point is to extract sufficient information about the target event in the embedding. A naive adaptation of MSM to distributed sensor systems does not differentiate between background and the intended event, resulting in embeddings with excessive background information. This concept is implemented by simultaneous learning of the sensor-masking independent embedding extraction and downstream tasks. Here, the learning of the downstream task acts as a GUIDE to ensure that the embedding contains sufficient information of the event.

To validate the proposed method in the DiMSEA, we recorded new datasets: MM-Store and MM-Office. Our datasets record human activity with multi-channel cameras and microphones distributed in convenience store and office environments. This concept differs from conventional multi-sensor datasets: datasets with non-distributed multi-view sensors~\cite{mvdataA1,mvdataA2,mmact,mvdataB1, mvdataB2,mvdataB3} or multi-modal datasets based on single point observations~\cite{avdataset1,avdataset2}. The experimental results on these datasets show that introducing Guided-MELD significantly improves the performance of DiMSEA, outperforming existing inter-sensor relationship modeling methods. It also performed robustly when reducing the number of sensors.

It should be noted that a part of this paper extends \textit{our conference paper}~\cite{multitrans}. The primary connections between the conference paper and this paper are as follows: (1) MultiTrans, the primary contribution of \textit{our conference paper}, serves as the encoder for the Guided-MELD that is newly proposed in this paper. The Guided-MELD is independently formulated as described in Sec.IV-B, thus representing a distinct contribution. Within this context, MultiTrans is clearly defined as a conventional method. As Sec. VI experimentally shows that Guided-MELD considerably improves the performance of DiMSEA beyond what MultiTrans achieves alone. (2) MM-Office has already been released in the \textit{our conference paper} among the two datasets we designed. The newly released MM-Store deals with a larger environment and more complex obstacles. To capture this environment, 1.5 times more sensors are deployed, creating a scenario better suited for DiMSEA.

\section{Related Studies}
\subsection{Conventional event analysis tasks}
An event analysis task using a single camera or microphone is a well-established problem setting~\cite{actionrec1, actionrec2, sed1, sed2}. Human action recognition, a fundamental research topic in computer vision, achieves high performance by extracting and classifying features using deep neural networks (DNN)~\cite{actionrec1,actionrec2, actionrec4}. Corresponding to such recognition tasks of computer vision, the task of detecting events that occur in the surrounding environment using sound as the input modality is called sound event detection~\cite{sed1,sed2}. Due to their narrow coverage, such single-sensor event detection methods have limitations in understanding human activity in an extensive real-world environment.

Several existing tasks use multiple sensors for event detection. These are broadly classified into two problem settings: (A) those in which multiple cameras are placed around a narrow target area ~\cite{mvcnn,danet,mmact,mva4,mva5} and (B) those in which multiple sensors are mounted on a self-driving machine ~\cite{crossview,deepfusion}, as shown in Figure~\ref{fig:multiview}. The main task belonging to setting (A) is a multi-view action recognition. This sensor arrangement aims at a detailed understanding of a target in a narrow area. On the other hand, sensor placement (B) aims to understand the environment surrounding a self-driving machine and is used for tasks such as passerby detection, lane detection, and mapping. In this case, multiple sensors are utilized to ensure that information about the machine's surroundings is not missed in all directions. In both cases, multiple sensors are densely utilized and are not used to cover an extensive area.

To cover an extensive area, a straightforward approach is to use distributed sensors, as shown in Figure~\ref{fig:mmstore}. Indeed, recent studies have shown that spatial features acquired with distributed microphones can be useful for understanding acoustic scenes~\cite{distmic}. Combining sensors of different modalities is also expected to be helpful~\cite{multi5, multi6, multi0, multi1,multi2, multi3, multi4}.

\subsection{Multi-sensor fusion}
\label{sec:RelatedStudy}
Multi-sensor fusion means the fusion of sensor observations from multiple viewpoints and modalities for various tasks. It includes multi-modal fusion, multi-view fusion, and both. 

Various multi-modal fusion methods have been proposed for extracting inter-modal relationship and fusing representations~\cite{multimodalfusion1,multimodalfusion2,multimodalfusion3,transfuser,multi3,multi5, multi6}. In particular, the use of multi-head self-attention (MHSA) based frameworks such as Transformer~\cite{attention,vit,asrtrans} have recently been reported to be effective for linking different modalities~\cite{videobert,transfuser,multi4}. MHSA is defined as the case where $Q=K=V$ in the following equation,
\begin{equation}
\label{eq:mhsa}
\begin{split}
\multihead(Q, K, V) = \concat(\head_1,\ldots,\head_h)W_O \\
{\rm where}\hspace{2pt}\head_i = \softmax\left(\frac{QW^Q_i(KW^K_i)^{\rm T}}{\sqrt{d_k}}\right)VW^V_i
\end{split}
\end{equation}
Here, $W^Q_i$, $W^K_i$, and $W^V_i$ are projection matrix, and $d_k$ are dimensions of $K$. ${\rm T}$ denotes the trace of the vector or matrix. 
Although MHSA was originally proposed for natural language processing (NPL), it has been reported that by using joint sequences from two modalities as input (e.g., video and word or image and LiDAR image), the relationship between them can be extracted~\cite{videobert,transfuser, multi4}.

On the other hand, there have been few studies on multi-view joint representation~\cite{mvcnn, danet, mmact, mva4, mva5, mvc2}. To the best of our knowledge, only Wang {\it et al.}~\cite{danet} have explicitly considered the inter-view relationship among cameras to analyze human action. In their work, a CRF is introduced to exchange messages among views~\cite{crfpaper}. Considering such an inter-view relationship enables multiple views to cooperate more effectively. The CRF can be expressed as the following iterative form by introducing the mean-field approximation,
\begin{equation}
 \label{eq:crf}
 \bm{z}_v^{(m)} = \frac{1}{\alpha_v}\left(\alpha_v \bm{z}_v^{(0)}+\sum_{u\neq v}\left(W_{u,v}\bm{z}_u^{(m-1)}\right)\right)
\end{equation}
where $z_v^{(m)}$ is the $m$-iteration of the feature of the $v-$th view. Learnable parameters of the model are $\alpha_v$, which is a coefficient to determine how much the features of the original view are maintained, and $W_{u,v}$, which is a coupling coefficient between the $u, v$th view. In the implementation of Li {\it et al.}~\cite{danet}, the number of iterations was set to 1. This setting is equivalent to accepting the linear transformation of all views except itself as the interaction term.

\subsection{Masked signal modeling based representation learning}
\label{sec:MSM}
Masked signal modeling (MSM) models the signal by reconstructing the masked signal~\cite{cmmasksd,dm2,mam,mim2,msm1,mlm1,mlm2,m2d}. Masked language modeling (MLM) with Transformer produced breakthroughs in sentence comprehension in NLP~\cite{mlm1,mlm2}. Subsequent studies have reported that MSM is also beneficial for understanding images and acoustic spectrograms~\cite{m2d,mim1,mim2,msm1}.

For a conventional method, we notably focus on Masked Modeling DUO (M2D)~\cite{m2d}, which introduces the concept of MSM to learn general representation for various acoustic tasks. Inputs of M2D are two spectrograms: the masked spectrogram and the rest. These are embedded using two networks called ``Online'' and ``Target'' and learned to maximize their agreement. The parameters of the Target are not updated by backpropagation, but the following exponential moving average (EMA) of the Online is used,
\begin{equation}
\label{eq:ema}
\zeta\leftarrow\ \rho\zeta + (1-\rho)\theta,
\end{equation}
where $\zeta$ is the parameter of the target network, $\theta$ is the parameter of the online network, and $\rho$ is the decay rate.

Such representation learning using the online and target networks was first proposed in image modeling methods based on data augmentation invariance~\cite{simclr,byol,msn,byol-a}. The difference between M2D and these methods is that signal modeling is based on masking or a combination of various data augmentation. Masking-based modeling overcomes the disadvantage that different choices of data augmentation methods can result in performance differences. Another promising MSM method is the masked auto-encoder (MAE), which takes a masked signal as input and performs representation learning by means of a task that reconstructs the input itself~\cite{mim1,mim2,msm1,mlm1,mlm2}.

While these MSM methods have been useful for modeling signals observed by a single sensor, modeling the relationship of signals across multiple sensors, which is the subject of this paper, has yet to be explored. Data augmentation methods are also difficult to use for the axes of a sensor series, i.e., to produce data captured by different locations and types of sensors. Therefore, modeling methods based on data augmentation universality cannot be used either. In contrast, modeling via reconstructing masked information in latent space, as used in M2D, is compatible with our problem setting.

\begin{figure}[tb!]
 \begin{center}
 \includegraphics[width=0.8\linewidth]{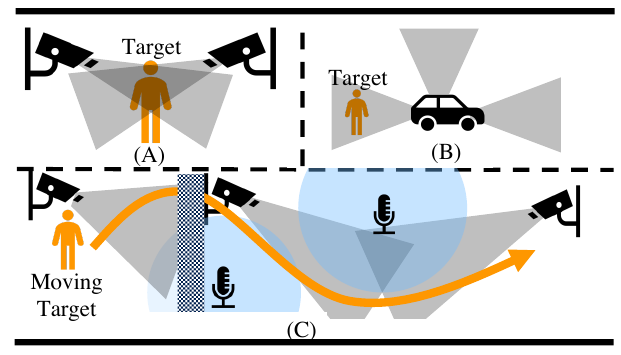}
 \caption{Variations of sensor placement for multi-sensor-based tasks.}
 \label{fig:multiview}
 \end{center}
 \end{figure}

\section{Distributed multimedia sensor event analysis (D\lowercase{i}MSEA)}
\label{sec:problemsetting}
\subsection{Formulation of multi-sensor-based event analysis}
In considering the distributed multimedia sensor event analysis (DiMSEA), we first formalize the general task of multi-sensor-based event analysis. Multi-sensor-based event analysis takes as input the observations $\{\bm{O}_{k,\tau}\}_{k\in\mathscr{K}_N}$ from $N$ sensors installed within a finite three-dimensional space $V$ to detect events occurring within $V$. Here, $\mathscr{K}_N=\{1,\ldots,N\}$ is the entire set of sensor indexes, $\tau\in\{1,\ldots,T\}$ is a discrete time index,  $\bm{O}_{k,\tau}$ is the signal observed at time $\tau$ by the $k$-th sensor. For the sake of generalized formulation, $\{\bm{O}_{k,\tau}\}_{k\in\mathscr{K}_N}$ shall include signals from any type of sensor.
An event is a semantic representation by classifying the situation inside $V$ at a certain time $\tau$ into a class $c\in\{1\ldots\ C\}$, determined through human interpretation. Here, $c$ denotes the index for the class of event, and $C$ is the number of considered classes. Considering that multiple events can occur simultaneously within the observation area $V$, the event at time $\tau$ can be represented by a multi-hot vector $\bm{y}_{\tau}=(y_{1,\tau},\ldots,y_{C,\tau})^{\rm T}\in\{0,1\}^{C}$.

The event detector outputs estimations of event activity $\hat{\bm{y}}_{\tau}$ at each time frame. A typical event detector utilizes an encoder $\mathcal{E}_{\theta}(\cdot)$, which extracts embedding vectors from sensor observations, and a classifier $\mathcal{C}_{\xi}(\cdot)$. Here, $\theta$ and $\xi$ are the parameters of the encoder and classifier, respectively. Given the observations $\{\bm{O}_{k,\tau}\}_{k\in\mathscr{K}_N}$, the posterior probability of the event $\hat{\bm{p}}_{\tau}=(\hat{p}_{1,\tau},\ldots,\hat{p}_{C,\tau})^{\rm T}$ is estimated as follows,
\begin{equation}
\label{eq:detection}
\hat{\bm{p}}_{\tau} = \mathcal{C}_{\xi}(\mathcal{E}_{\theta}(\{\bm{O}_{k,\tau}\}_{k\in\mathscr{K}_N}))
\end{equation}
The outputs of event detection $\hat{\bm{y}}_{\tau}$ are obtained by applying an appropriate threshold $\alpha\in[0,1)$ to the obtained posterior probabilities. 

The optimal parameter can be obtained, for example, by minimizing binary cross-entropy between the ground-truth labels and predictions,
\begin{equation}
\label{eq:strong-bce}
\underset{\theta,\xi} {\operatorname{min}} -\frac{1}{C}\left[\sum_{c=1}^C y_{c,\tau}\log\hat{p}_{c,t} + (1 - y_{c,\tau})\log(1-\hat{p}_{c,\tau})\right]
\end{equation}
As addressed in this paper, when training the event detector with weak labels, which are given only as the class of events occurring within a certain time interval, the event detector cannot possibly be directly trained with Eq.~\ref{eq:strong-bce}. Instead, the event detector can be indirectly trained by multiple instance learning~\cite{mil} that minimize the binary cross-entropy between the marginalized outputs of event detector $\hat{\bm{p}}^{\mbox{\scriptsize bag}}$ over time and weak labels $\bm{y}^{\mbox{\scriptsize bag}}$,
\begin{equation}
\label{eq:weak-bce}
\underset{\theta,\xi} {\operatorname{min}} -\frac{1}{C}\left[\sum_{c=1}^C y^{\mbox{\scriptsize bag}}_{c}\log\hat{p}^{\mbox{\scriptsize bag}}_{c} + (1 - y^{\mbox{\scriptsize bag}}_{c})\log(1-\hat{p}^{\mbox{\scriptsize bag}}_{c})\right],
\end{equation}
where, the weak label $\bm{y}^{\mbox{\scriptsize bag}}$ is defined as
\begin{equation}
y^{\mbox{\scriptsize bag}}_c = 
 \begin{cases}
 1\hspace{10pt}{\rm if}\hspace{10pt}^{\exists}\tau,\, y_{c,\tau}=1\\
 0\hspace{10pt}{\rm if}\hspace{10pt}^{\forall}\tau,\,y_{c,\tau}=0
 \end{cases}
\end{equation}

\subsection{Formulation of DiMSEA}
\label{sec:dsea}
We aim to detect human behavior in extensive and complex environments like stores. For this purpose, multiple sensors need to be used for two reasons. First, in complex environments, information from a single sensor is fragmentary, necessitating cooperation among multiple sensors to identify events. Second, to cover extensive areas, the cooperation of multiple sensors is also required. Therefore, for our purpose, it is essential to utilize an integrated distributed sensor system, as shown in Fig.~\ref{fig:multiview} (C), where each sensor covers different areas with partially overlapping coverage. Distributed multimedia sensor event analysis (DiMSEA) is a task we have newly designed to analyze events in extensive and complex environments by coordinating such a distributed sensor system.

In information theory, the reduction in uncertainty about an event $\bm{y}_{\tau}$ given sensor observations is described as the mutual information between them.
The mutual information between the sensor system's observations $\{\bm{O}_{k,{\tau}}\}_{k\in\mathscr{K}_N}$ and the event $\bm{y}_{\tau}$ is expressed as follows,
\begin{equation}
\label{eq:MI1}
I_{\mathscr{K}_N; \bm{y}_\tau} = I(\{\bm{O}_{k,\tau}\}_{k\in\mathscr{K}_N};\bm{y}_{\tau})
\end{equation}
Considering an appropriate system for estimating the event $\bm{y}_{\tau}$ from observations $\{\bm{O}_{k,\tau}\}_{k\in\mathscr{K}_N}$, a greater $I_{\mathscr{K}_N; \bm{y}_\tau}$ indicates reduced uncertainty in the estimation. Problem setups (A)-(C) are all designed so that the sensor system provides sufficient information $I_{\mathscr{K}_N; \bm{y}_t}$ to detect the target event in each scenario.

It is considered that the contribution of a sensor $\kappa\in\mathscr{K}_N$ in the entire sensor system to event analysis depends on two factors. The first is the amount of information about the event observed by sensor $\kappa$, denoted as $I_{\{\kappa\},\bm{y}_{\tau}}$. A sensor with $I_{\{\kappa\},\bm{y}_{\tau}}\simeq0$ does not provide information about the event, potentially acting as noise in the system. Such sensors, called background sensors, are installed to extend the coverage of the sensor system and are expected to provide useful information at different time frame. The second factor is the reduction of the uncertainty of the event $\bm{y}_{\tau}$ when the observation of sensor $\kappa$ is added under the condition that the observations of other sensors are already given. It is formulated as the conditional mutual information:
\begin{align}
\delta I_{\kappa; \bm{y}_{\tau}}
&= I(\bm{O}_{\kappa,\tau};\bm{y}_{\tau}|\{\bm{O}_{k,\tau}\}_{k\in\mathscr{K}_N\setminus\{\kappa\}})\\
&= I_{\mathscr{K}_N; \bm{y}_{\tau}} - I_{\mathscr{K}_N\setminus\{\kappa\}; \bm{y}_{\tau}}
\end{align}
A sensor $\kappa$ with a high $\delta I_{\kappa; \bm{y}_{\tau}}$ is considered to provide unique information to the system. Hence, even if the information $I_{\{\kappa\},\bm{y}_{\tau}}$ it holds about the event is small, it is deemed valuable. Conversely, a sensor $\kappa$ with a low $\delta I_{\kappa; \bm{y}_{\tau}}$ contributes little new information to the entire sensor system. Even if sensor $\kappa$ has information about the event, i.e., $I_{\{\kappa\},\bm{y}_{\tau}}$ is significantly large, the information is redundant if it is equivalent to information already observed by other sensors.

The multi-sensor-based problem settings shown in Fig.~\ref{fig:multiview} differ in how each sensor contributes to the sensor system. (A) does not include background sensors, and each sensor provides redundant or unique information. (B) consists mostly of background sensors, with one or a few providing unique information about the event. The problem setting for DiMSEA (C) includes all cases above. This sensor system includes background and non-background sensors, the number of which varies depending on the situation. Moreover, it contains both redundant and unique information. Therefore, a system for DiMSEA must discern the necessary sensors in accordance with the situation and effectively extract complementary information from the redundant information contained therein. This is a different problem setting from existing multi-sensor-based recognition and detection tasks.

\section{Guided Masked self-distillation modeling (Guided-MELD)}
To address the challenges in the DiMSEA, we propose {\bf Guided} {\bf M}asked s{\bf EL}f-{\bf D}istillation modeling (Guided-MELD). This method extracts a joint representation of distributed sensor observation that distills relevant information about the target event from observations of distributed sensors while ignoring obstructive information. This section clarifies the difficulties in the DiMSEA and presents the general formulation of the proposed Guided-MELD approach and its implementation.

\subsection{Major difficulty in DiMSEA}
As discussed in Sec.~\ref{sec:problemsetting}, the DiMSEA system need to be able to focus attention on the sensors required by the situation and to extract unique and complementary information from their redundant observations. This integration of sensor information, i.e., the modeling of inter-sensor relationships, is conducted through an encoder \(\mathcal{E}_{\theta}\) that uses observations from multiple sensors or their features as inputs. This section discusses the properties that encoder \(\mathcal{E}_{\theta}\) should possess for DiMSEA and the difficulty in developing such an encoder.

\begin{figure}[t!]
 \centering
 \includegraphics[width=0.7\linewidth]{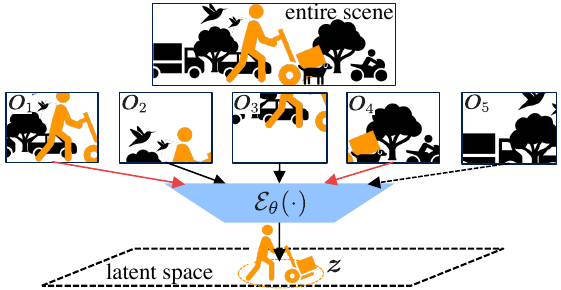}
 \caption{A schematic view of a desirable encoder for DiMSEA. In this example, the entire scene is observed by five distributed sensors. The time index is omitted here for simplicity.}
 \label{fig:gmsmconcept}
\end{figure}

Fig.~\ref{fig:gmsmconcept} schematically illustrates the desirable encoder \(\mathcal{E}_{\theta}\) for DiMSEA. In this situation, Sensor 0 observes the most information about the event. Sensors 0, 1, and 2 are redundant with each other. Sensor 3, despite containing much noise, is used because it holds unique information about the event. Sensor 4, a background sensor, can potentially degrade system performance. In contrast, once sensor 0 becomes unavailable due to obstacles, the information from sensors 1, 2, and 3 becomes unique in this scenario. The desirable encoder \(\mathcal{E}_{\theta}\) should understand such situations and semantically select information from each sensor to embed only event-related information in the latent space.

A gap remains between conventional methods of modeling inter-sensor relationships and this desirable encoder. In CRF (Eq.~\ref{eq:crf}), sensor relationships are modeled using a fixed weight matrix. An encoder based on CRF has high weights for averagely valuable sensors. Such a system degrades in performance in the situation where the focused sensor becomes unavailable. Our prior study addresses this issue with the MultiTrans, which can pay attention to necessary sensors depending on input sensor signals. However, even with such input-dependent attention mechanisms, the issue of the over-reliance on specific sensors persists. This can be attributed to the huge variety of situations and sensor combinations in DiMSEA settings, making it challenging to learn these scenarios through training comprehensively. Indeed, our experimental results show that the performance of systems trained with CRF or MultiTrans is significantly degraded by removing only a few sensors.

As an alternative approach for modeling the relationships between sensors, it is conceivable to utilize Masked Signal Modeling (MSM), which has been successful in modeling various modalities of signals for modeling inter-sensor relationships. Although no existing studies exist, one naive implementation could involve masking some sensor signals and training an encoder to restore them. Since this method aims to model inter-sensor relationships by restoring missing sensor information, the encoder learned through this process is expected to depend less on any specific sensor. However, unlike cases where MSM has been successful, the signals from distributed multimedia sensors are mostly background signals. Since MSM itself does not differentiate between background and target events, such a naive adaptation of MSM might not result in embeddings that sufficiently contain information about events. Indeed, previous studies have reported embedding noisy information in MSM, leading to degraded task performance~\cite{ask2mask,attnmsm}.

\begin{figure*}[t!]
 \centering
 \includegraphics[width=0.88\linewidth]{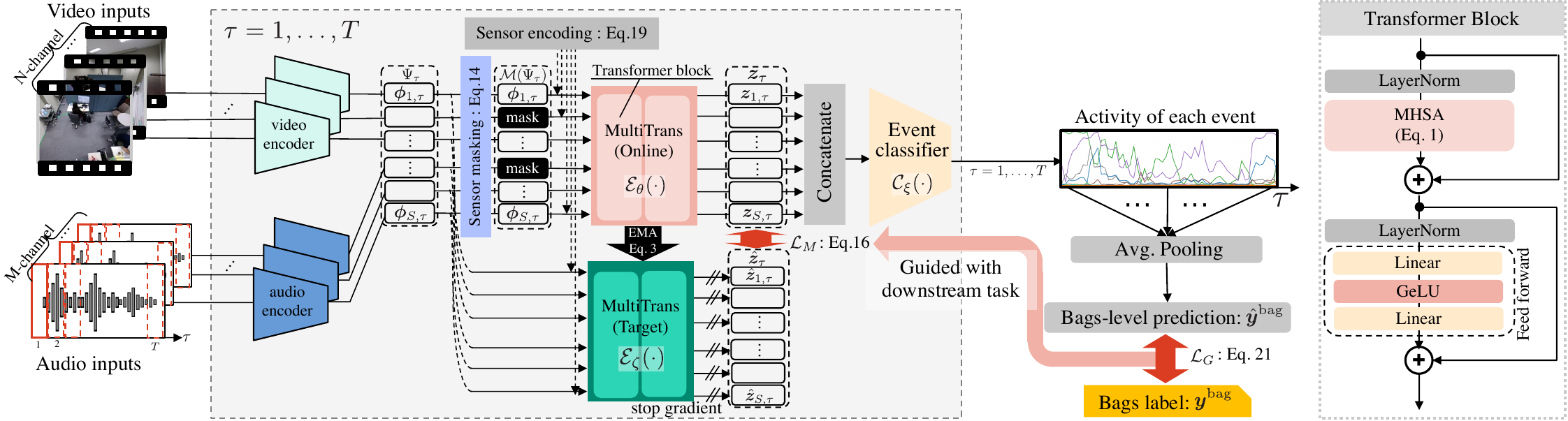}
 \caption{The architecture of Guided-MELD for distributed multi-modal event detection tasks.\\{\footnotesize The N-channel camera and M-channel microphone input signals are divided into $T$-frames ($\tau$ is the index of time frame), and the event classifier output $\hat{\bm{p}}_{\tau}$ is computed in parallel. See sec.~\ref{sec:impl} for details of each block. For more details on video and audio input shaping, see Fig.\ref{fig:inputshaping}.}}
 \label{fig:architecture}
\end{figure*}

\subsection{General formulation of Guided-MELD}
Guided MELD is designed for extracting embeddings that contain sufficient information on target events from redundant, fragmented, or noisy distributed multimedia sensor signals without over-reliance on specific sensors.
The principle of the proposed method is to train the encoder to embed as much event-related information as possible in the latent representation while any arbitrary sensor set is missing. 
Here, assuming an ideal embedding $\bm{z}^*_{\tau}\in\mathbb{R}^D$  extracted by an ideal encoder, $\bm{z}^*$  maximizes the performance of the event-related downstream tasks.
Given this ideal $\bm{z}^*$, the Guided-MELD principle can be expressed as follows,
\begin{equation}
\label{eq:gmsm}
\underset{\theta} {\operatorname{Minimize}}\hspace{3pt}\mathbb{E}_{\mathscr{K}\subset\mathscr{K}_N}\left[d(\mathcal{E}_{\theta}(\{ \bm{O}_{k,\tau}\}_{k\in\mathscr{K}}),\bm{z}^*_{\tau})\right],
\end{equation}
where, $d(\cdot,\cdot)$ denotes an arbitrary distance, e.g., $L_2$ distance.
By minimizing Eq.~\ref{eq:gmsm}, the encoder $\mathcal{E}_{\theta}$ is trained to extract embeddings that are consistent for the presence or absence of a particular sensor set and has sufficient information to solve downstream tasks. 

In practice, the ideal $\bm{z}^*_{\tau}$ cannot be prepared in advance for training. Therefore, Guided-MELD simultaneously maximizes the performance of downstream tasks and minimizes Eq.~\ref{eq:gmsm} to obtain an embedding $\bm{z}_{\tau}$ that asymptotically approaches $\bm{z}^*_{\tau}$.
This performance maximization of the downstream task is implemented as the following minimization of its cost function $\mathcal{L}_G$,
\begin{equation}
\label{eq:gmsm2}
\thickmuskip=0mu
\medmuskip=0mu
\thinmuskip=0mu
\underset{\theta, \xi} {\operatorname{Minimize}}\hspace{3pt}
\mathbb{E}_{\mathscr{K}\subset\mathscr{K}_N}\left[\mathbb{E}_{\tau}\left[\mathcal{L}_G(\mathcal{C}_{\xi}(\mathcal{E}_{\theta}(\{\bm{O}_{k,{\tau}}\}_{k\in \mathscr{K}})), \bm{y}_{\tau})\right]\right]
\end{equation}
In this joint optimization, the following $\hat{\bm{z}}_{\tau}$ is used instead for the target $\bm{z}^*_{\tau}$ in Eq.~\ref{eq:gmsm},
\begin{equation}
\label{eq:zhat}
\hat{\bm{z}}_{\tau} = \mathcal{E}_{\zeta}(\{\bm{O}_{k,{\tau}}\}_{k\in \mathscr{K}_N}),
\end{equation}
where $\mathcal{E}_{\zeta}$ is the mean teacher of $\mathcal{E}_{\theta}$ (See Eq.~\ref{eq:ema}).
Unlike the naive adaptation of MSM, since this $\hat{\bm{z}}_{\tau}$ is also expected to approach $\bm{z}^*_{\tau}$ through training, Eq.~\ref{eq:gmsm} is expected to be asymptotically satisfied at the end of training.
In other words, the modeling is GUIDED by maximizing performance on the downstream task. 

Note that the above formulation does not limit the encoder $\mathcal{E}_{\theta}$ architecture for Guided-MELD.  As mentioned below, MultiTrans is adopted for the encoder architecture as one implementation.

\subsection{Implementation for DiMSEA}
\label{sec:impl}
This section describes the implementation of the DiMSEA in Guided-MELD.  Fig.~\ref{fig:architecture} shows the network architecture of our proposed method implemented for DiMSEA. In Eq.~\ref{eq:gmsm}, embeddings are extracted from a subset of sensors through the encoder $\mathcal{E}_{\theta}$. This sensor subset selection is performed by applying a random mask to the extracted features of each sensor.  Although applying the mask directly to $\bm{O}_{k,\tau}$ is an option, this implementation did not use this method to avoid unnecessary learning by masked signals to the learnable encoders used for feature extraction for each sensor. The masking operation $\mathcal{M}(\cdot)$ on the $D-$dimensional sensor features $\Psi_{\tau}=(\bm{\phi}_{1,\tau},\ldots,\bm{\phi}_{S,\tau})\in\mathbb{R}^{D\times S}$ is expressed as
\begin{equation}
\label{eq:mask}
\mathcal{M}(\Psi_{\tau}) = \Psi_{\tau}\odot(\bm{\mu}_1,\ldots,\bm{\mu}_S).
\end{equation}

Here, $\odot$ denotes the Hadamard product. The $\bm{\mu}_s\in\mathbb{R}^{D}$ is either a zero vector $\bm{0}=(0,\ldots,0)^{\rm T}$ or a one vector $\bm{1}=(1,\ldots,1)^{\rm T}$, randomly chosen in accordance with the mask ratio $\kappa\in[0,1]$. This mask ratio $\kappa$ is a parameter that determines the ratio of sensor-masking, constraining the mask matrix $(\bm{\mu}_1,\ldots,\bm{\mu}_S)$ to satisfy the following condition,

\begin{equation}
\label{eq:mask2}
\sum_{s=1}^{S}\bm{\mu}_s=\lfloor (1-\kappa) S\rfloor\bm{1}
\end{equation}

Guided-MELD uses two networks termed "Online" and "Target". The parameters of the Target network are updated using Eq.~\ref{eq:ema}, which is the EMA of the Online network's parameters. In both of networks, the inputs consist of sequences of masked and unmasked sensor features, respectively. The outputs are the joint embeddings of these sensor features, represented as $\bm{z}_{\tau}=(\bm{z}_{1,\tau},\ldots,\bm{z}_{S,\tau})$ and $\hat{\bm{z}}_{\tau}=(\hat{\bm{z}}_{1,\tau},\ldots,\hat{\bm{z}}_{S,\tau})$ for the online and target networks, respectively. Using these, the minimization of Eq.~(\ref{eq:gmsm}) was implemented as the minimization of the following loss function,
\begin{equation}
\label{eq:loss}
\mathcal{L}_{\mbox{\scriptsize M}}
=\frac{1}{T}\frac{1}{\lfloor(1-\kappa)S\rfloor}\sum_{\tau=1}^{T}\sum_{s=1}^{S}\|(\hat{\bm{z}}_{s,\tau} - \bm{z}_{s,\tau})\odot(\bm{1} - \bm{\mu}_s)\|_2^{2}
\end{equation}
Here, $\|\cdot\|_2^2$ denotes the square of the $L_2$ norm. By multiplying with $\bm{1}-\bm{\mu}_{s}$, $\mathcal{L}_{\mbox{\scriptsize M}}$ focuses on maximizing agreement between $\bm{z}_{s,\tau}$ and $\hat{\bm{z}}_{s,\tau}$ for masked sensors.

The joint optimization of Eq.~\ref{eq:gmsm} and \ref{eq:gmsm2} was implemented as multitask learning by using the following loss function,

\begin{equation}
\label{eq:multitask}
\mathcal{L} = \mathcal{L}_{G} + \lambda \mathcal{L}_{M}
\end{equation}

Here, $\lambda$ is a balancing parameter. A high $\lambda$ at the beginning of training can make an encoder unsuitable for the task, whereas a small $\lambda$ throughout the training can diminish the effect of Guided-MELD. Therefore, $\lambda$ is scheduled as follows,
\begin{equation}
\label{eq:schedule}
\lambda = \lambda_0\gamma^n\hspace{5pt}(\gamma\geq1.0),
\end{equation}
In this equation, $n$ represents the number of training epochs, $\lambda_0$ is the initial value of $\lambda$, and $\gamma$ is a hyperparameter controlling the increase of $\lambda$.
\subsection{Implementation details}
\label{sec:detail}
The input of the system is synchronized $M$-channel videos and $N$-channel audio signals. As in existing works using multiple sensors~\cite{mvcnn,danet,mmact,crossview}, the video encoder $\mathcal{V}_e$ and shared audio encoder $\mathcal{A}_e$ are shared for all channels. The video encoder $\mathcal{V}_e$ is implemented as ResNet-34~\cite{resnet34} pre-trained on ImageNet, minus the output layer. In the audio encoder $\mathcal{A}_e$, the input audio is first transformed into the log-absolute value of the short-time Fourier transform (STFT) spectrogram, and then a Mel-filter bank is applied. The audio embeddings are extracted using VGGish~\cite{vggish} pre-trained with AudioSet. The extracted audio and video features are embedded in the $D$-dimensional vectors with a linear layer. For the obtained embedding $\psi$, sensor-masking is applied by Eq.~\ref{eq:mask}. The mask rate $P_S$, which can be determined independently for each sensor, is a common value for the camera and microphone, respectively. The masked embedding $\psi_m$ is input to MultiTrans (Online) after applying the following sensor encoding,
\begin{equation}
 \tilde{\bm{\phi}}_{s}= \concat(\bm{\phi}_{s}, \onehot_S(s)) \in \mathbb{R}^{D+S}.
 \label{eq:viewenc}
\end{equation}
On the other hand, the input to MultiTrans (Target) is the same sensor encoding applied to the unmasked embedding $\psi$. Both Online and Target of MultiTrans have the same structure, which is a stack of multiple Transformer blocks consisting of $H$-head MHSA.

The classifier $\mathcal{C}_{\xi}$ for estimating event activity $\hat{\bm{p}}_{\tau}$ from the fused feature is implemented as a linear layer with a sigmoid activation function. The activation of the event $\hat{\bm{y}}_{\tau}$, the output of the system, is 1 if $\hat{\bm{p}}_{\tau}$ exceeds a fixed threshold and 0 otherwise.

The whole system is trained using the weak label $\bm{y}^{\mbox{\scriptsize bag}}$ in a multiple-instance learning (MIL) manner~\cite{mil}, often used in weakly supervised sound event detection~\cite{sed3}. In the MIL scheme, the following bags-level prediction $\hat{\bm{p}}^{\mbox{\scriptsize bag}}$ is first calculated from the obtained event activity sequence as,
\begin{equation}
\label{eq:bagpred}
\hat{\bm{p}}^{\mbox{\scriptsize bag}} = \frac{1}{T}\sum_{\tau=1}^{T}\hat{\bm{p}}_{\tau}
\end{equation}
Since the dataset we used has a large class imbalance, we use the following weighted BCE~\cite{wbce},
\begin{equation}
\thickmuskip=0mu
\medmuskip=0mu
\thinmuskip=0mu
\mathcal{L}_{\mbox{\scriptsize{G}}}=-\sum_{c=1}^{C}
w_c\left(y^{\mbox{\scriptsize bag}}_{c}\log(\hat{p}^{\mbox{\scriptsize bag}}_{c}) \right.
\left. +(1-y^{\mbox{\scriptsize bag}}_{c})\log(1-\hat{p}^{\mbox{\scriptsize bag}}_{c}) \right)
\end{equation}
where $w_c$ is the reciprocal of the total number of events in a dataset belonging to the $c-$th class.

Parameters for MultiTrans (Target) in Guided-MELD were updated by applying Eq.~\ref{eq:ema} at every iteration. Note that stop gradient was introduced here to avoid updating the MultiTrans (Target) parameters via backpropagation.

\section{Proposed datasets}
We collected new datasets called MM-Store and MM-Office. In this section, we describe their details.

\subsection{MM-Store}
\label{sec:mmstore}
MM-Store was recorded in the convenience store shown in Figure ~\ref{fig:mmstore}. Twelve omnidirectional microphones and six cameras, shown in Table~\ref{tb:sensorspec}, were used to record audio and video. The amount of data was 990 clips per location and sensor, divided into 792, 66, and 132 clips each for training, validating, and testing. Each clip, ranging from 30 to 90 seconds, contained two to five events selected from 18 classes of events. Each event is designed to simulate an action in a convenience store and is performed by one clerk and one customer each. The labels for training were given as multi-labels indicating what events each clip contained. For evaluation, only the test data was given a strong label containing the onset/offset time of each event.

\begin{table}[t!]

\caption{Specifications of sensors used to record MM-Store and MM-Office datasets.}
\label{tb:sensorspec}
\centering
\scalebox{0.9}[0.9]{
\begin{tabular}{ll|cc}

\toprule
 & Dataset & MM-Store & MM-Office \\ \midrule
\multirow{3}{*}{Camera} & Model number & ARUCOM RD-CI502 & GoPro Hero8 \\
 & Resolution & $3840\times2160$ & $2560\times1440$ \\
 & FPS & 20 & 30 \\ \midrule
\multirow{4}{*}{Microphone} & Model number & \multicolumn{2}{c}{HOSIDEN KUB4225} \\
 & Directivity & \multicolumn{2}{c}{Omni-directional} \\
 & Bit rate & \multicolumn{2}{c}{32 bits} \\
 & Sampling rate & \multicolumn{2}{c}{48kHz} \\ \bottomrule
\end{tabular}
}
\end{table}

\begin{figure}[tb!]
 \begin{center}
 \includegraphics[width=0.77\linewidth]{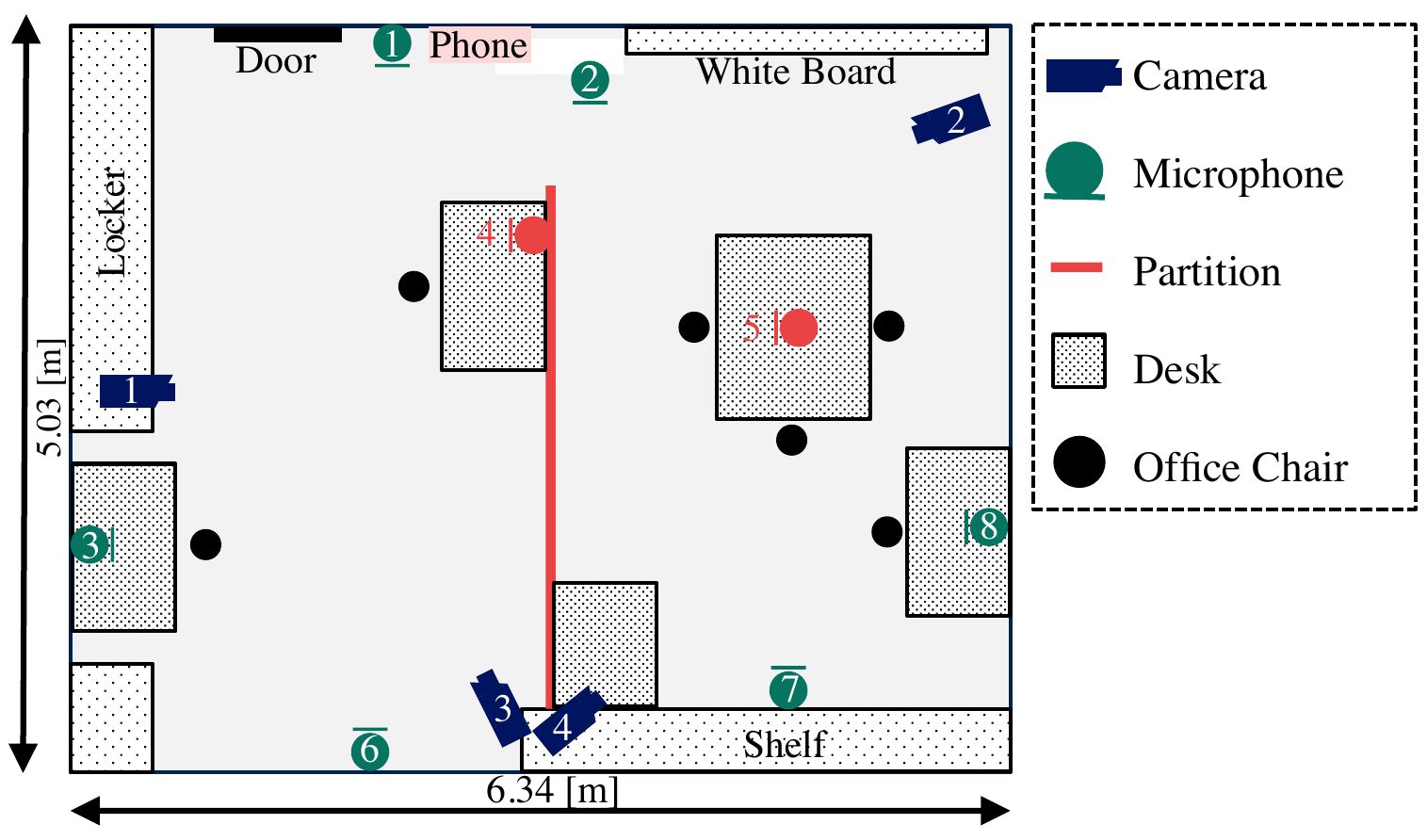}
 \caption{Room and sensor setup of MM-Office dataset}
 \label{fig:roomsetup}
 \end{center}
 \end{figure}
\subsection{MM-Office}
\label{sec:mmoffice}
MM-Store was recorded in an office room as shown in Figure ~\ref{fig:roomsetup}. Eight omni-directional microphones and four cameras, shown in Table~\ref{tb:sensorspec}, were used to record audio and video. The data volume was 880 clips per location and sensor, divided into 704, 88, and 88 clips each for training, validating, and testing. Each clip, ranging from 30 to 90 seconds, contained one to four events selected from 12 different classes of events (see \cite{multitrans} for details). Each event is intended for daily office work (e.g. enter office, meeting and eat) and is performed by one to three persons. 
\section{Experiments}
\subsection{Comparison method}
Comparative experiments are conducted to evaluate the effectiveness of the proposed method in the DiMSEA task. To the best of our knowledge, no existing study has addressed the event analysis task utilizing distributed multimedia sensors. As a similar field, we refer to action recognition with multiple cameras. In particular, we have chosen CRF-based modeling~\cite{danet} and MultiTrans-based modeling~\cite{multitrans} as comparison methods, as they are the state-of-the-art methods for inter-view relationship modeling. As baseline methods, we set up a "Late fusion" method, which uses each sensor independently to fuse only the results, and a "Concat" method, which simply combines the features obtained from each sensor. In addition to these end-to-end learning-based methods, we employed a method based on MSM as a comparison method. The M2D~\cite{m2d} was selected as the state-of-the-art method for MSM in the feature domain. 

Although various methods can be considered for the model to extract the feature from the input of each sensor, the same condition for feature extraction was used in all the methods to focus on the modeling of inter-sensor relationships in this comparative experiment. The use of more advanced encoders for video~\cite{videorep1,videorep2} and audio~\cite{byol-a,m2d} and the incorporation of multi-modal correlations at the encoding stage~\cite{multimodalfusion2,transfuser} are considered to contribute to performance improvement, but these factors are considered to be orthogonal to the inter-sensor relationship modeling.The detailed implementation of each comparison method is described below.

\begin{description}
\item{\bf (A) Late Fusion} This method detects events independently at each sensor and fuses the results. The average and maximum are considered as simple fusion methods, and the average was adopted here because of its better performance. It is expressed as following,
\begin{equation}
 \hat{\bm{p}}_{\tau} = \frac{1}{S}\sum_{s=1}^S\mathcal{C}_s(\bm{\phi}_{s,\tau}),
\end{equation}
where $\mathcal{C}_s$ denotes the event classifier for the s-th sensor. By comparing it with the proposed method, we can clarify the difference in performance due to cooperation between sensors.

\item{\bf (B) Concat} This method uses a simple concatenation of each sensor's features as fused features for event detection. It is expressed by the following equation,
\begin{equation}
 \hat{\bm{p}}_{\tau} =\mathcal{C}_{\xi}\left(\concat(\bm{\phi}_{1.\tau},\ldots,\bm{\phi}_{S,\tau})\right)
\end{equation}
By comparing it with the proposed method, we can clarify the performance difference due to considering the inter-sensor relationship in the feature fusion stage.

\item{\bf (C) CRF}~\cite{danet} This method uses the fused features from each sensor using CRFs to perform event detection. It is expressed by the following equation,
\begin{equation}
 \hat{\bm{p}}_{\tau} =\mathcal{C}_{\xi}\left(\concat({\rm CRF}_1(\bm{\phi}_{1.\tau},\ldots,\bm{\phi}_{S,\tau}))\right),
\end{equation}
where ${\rm CRF}_1$ denotes the 1st iteration of Eq.~\ref{eq:crf} as used in DANet~\cite{danet}. Comparing it with the proposed method reveals performance differences due to introducing Guided-MELD, i.e., the flexibility to change attention to each sensor depending on the input.

\item{\bf (D) MultiTrans}~\cite{multitrans} This method uses the fused features from each sensor using MultiTrans to perform event detection. It is expressed by the following equation,
\begin{equation}
\thickmuskip=0mu
\medmuskip=0mu
\thinmuskip=0mu
 \hat{\bm{p}}_{\tau} =\mathcal{C}_{\xi}\left(\concat({\rm MultiTrans}(\tilde{\bm{\phi}}_{1.\tau},\ldots,\tilde{\bm{\phi}}_{S,\tau}))\right).
\end{equation}
By comparing it with the proposed method, we can clarify the effects of the proposed joint optimization of Eq.~\ref{eq:gmsm} and Eq.~\ref{eq:gmsm2}.

\item{\bf (E) MSM (M2D)}~\cite{m2d} This method is a naive adaptation of M2D-like masked signal modeling methods to make them able to model distributed sensor signals. This model is trained in two steps. First, the encoder $\mathcal{E}_{\theta}$ is learned by minimizing the loss function $\mathcal{L}_M$ in Eq.~\ref{eq:gmsm}. 
Note that, following the M2D implementation, the mask is applied in Target's network in reverse to that for Online.
Next, by freezing the obtained parameters, only the classifier $\mathcal{C}_{\xi}$ is trained by minimizing the loss function $\mathcal{L}_G$ in Eq.~\ref{eq:gmsm2}. By comparing this with the proposed method, the effectiveness of guiding joint representation learning with the downstream task.

\item{\bf (F) Ours: Guided-MELD} The proposed method utilizes Guided-MELD.
\end{description}

In addition to these five methods, we evaluated two partially modified or degraded methods from the proposed method as an ablation study.

\begin{description}
\item{\bf (G) Guided-MELD w/o $\mathcal{L}_{M}$} This method is Guided-MELD without $\mathcal{L}_{M}$ Eq.~\ref{eq:gmsm}. That is, it is (D) plus sensor-masking only. By comparing this method with the proposed method, it is possible to clarify whether the performance difference from (D) to (F) is due to the masking of the sensor information itself or not.

\item{\bf (H) Guided-MELD (CRF)} This method replaces MultiTrans, which is used for multi-sensor fusion in Guided-MELD, with CRF. By comparing this method with the proposed method, it is possible to examine whether the proposed method is effective in combination with MultiTrans or if it is effective when modeling between sensors using another method.
\end{description}

\subsection{Experimental procedure}
To validate the effectiveness of the proposed method from multiple perspectives, four experiments were conducted using the MM-Store dataset. All experiments were conducted three times with different random seeds to account for the effect of initial values of network parameters.

\begin{description}

\item{\bf Event tagging experiment} Event tagging is a subtask of DiMSEA. This task identifies all classes of the event in a clip. It can also be an event detection task without specifying when an event occurs or ends. By evaluating the performance of this task, the performance of each system can be independently evaluated in terms of its ability to identify the class of event.
 
\item{\bf Event detection experiment} Event detection is a subtask of DiMSEA. This task identifies the class of event and its onset/offset time. Our dataset contains up to two simultaneous events, which need to be able to be detected separately.

\item{\bf Sensor reduction experiment} We conducted sensor reduction experiments to evaluate the ability of Guided-MELD to compensate for missing sensor information. The experiment evaluates the performance of event detection when one or two sensors are masked. The detection model utilizes a model that has been trained with all sensors without having to re-train it. For a masking method, we used a method similar to Eq.~\ref{eq:mask}. The numbers for reducing 1 and 2 sensors are ${}_{18}\mathrm{C}_1 = 18, {}_{18}\mathrm{C}_2 = 153$, respectively, and we performed inferences using all these combinations.

\item{\bf Investigation of guiding schedule} We investigated scheduling strategies for two task weights $\lambda$ in the loss function Eq.~\ref{eq:multitask} introduced in Guided-MELD. Comparative experiments were conducted on three scheduling strategies.
\begin{description}
\item {{\bf(F-1)} Decrease $\lambda$:} We use the following scheduling instead of Eq.~\ref{eq:schedule},
\begin{equation}
 \lambda = \lambda_{\mbox{\scriptsize{max}}}\gamma^{-n}
\end{equation}
Here, $\lambda_{\mbox{\scriptsize{max}}}$ is the value of $\lambda$ at the maximum number of epochs of learning in Eq.~\ref{eq:schedule}. The $n$ is the number of epochs of training. This scheduling strategy emphasizes an event analysis task in the late stages of learning.

\item {{\bf (F-2)} Constant $\lambda$:} Instead of using Eq.~\ref{eq:schedule}, we use a fixed weight $\lambda=\bar{\lambda}$. The $\bar{\lambda}$ is determined as in the following equation,
\begin{equation}
\bar{\lambda} = \frac{\sum_{n=0}^{\scriptsize{\rm MaxEpoch}}\lambda_0 \gamma^n}{\rm MaxEpoch}
\end{equation}

\item{{\bf (F-3)} Increase $\lambda$:} $\lambda$ is varied in accordance with Eq.~\ref{eq:schedule}.
\end{description}
\end{description}

In addition to the above, to validate the proposed method using a different dataset, experiments 1 and 2 were performed even when using the MM-Office dataset.

\subsection{Evaluation metrics}
We adopted the mean average precision (mAP) and receiver operating characteristic area under the curve (ROAUC) as evaluation metrics for the distributed multimedia sensor event analysis task, which are widely adopted in detection tasks. The average precision (AP) is the area under the precision-recall curve. The mAP is obtained as the mean of AP over classes. Note that there are two ways of averaging APs: macro mAP, which calculates mAPs for each class and then averages them, and micro mAP, which averages all classes at once. In this experiment, we adopt macro mAP. Higher values of mAP indicate better performance. The ROAUC is the area under the ROC curve that plots the true positive rate (TPR) versus the false positive rate (FPR)~\cite{auc}. Higher values of ROAUC indicate better performance. Random inference results in an ROAUC of 0.5. The TP, FP, TN, and FN for calculating these metrics are counted over all test data frame-by-frame for the event detection task and clip-by-clip for the event tagging task.

\begin{table}[t!]
\caption{List of hyperparameters used in our experiments}
\label{tb:hyperparameter}
\scalebox{0.86}[0.86]{
\begin{tabular}{@{}llcc@{}}
\toprule
 & \multicolumn{1}{l|}{Dataset} & MM-Store & MM-Office \\ \midrule
System & \multicolumn{1}{l|}{Camera} & $M=6$ & $M=4$ \\
input& \multicolumn{1}{l|}{Microphone} & $N=12$ & $N=8$ \\ \midrule
Input & \multicolumn{1}{l|}{Input length} & \multicolumn{2}{c}{$25.6sec$} \\
length & \multicolumn{1}{l|}{Time frames} & \multicolumn{2}{c}{$T=32$} \\ \midrule
Video & \multicolumn{1}{l|}{Video resolution} & \multicolumn{2}{c}{$224\times224$} \\
pre-process. & \multicolumn{1}{l|}{Standardization} & \multicolumn{2}{c}{ImageNet's mean and std. values} \\ \midrule
 & \multicolumn{1}{l|}{Down sampling} & \multicolumn{2}{c}{16kHz} \\
Audio & \multicolumn{1}{l|}{FFT length} & \multicolumn{2}{c}{400} \\ 
pre-process. & \multicolumn{1}{l|}{hop size} & \multicolumn{2}{c}{160} \\ 
 & \multicolumn{1}{l|}{Mel filterbank} & \multicolumn{2}{c}{64} \\ \midrule
Video & \multicolumn{1}{l|}{Model} & \multicolumn{2}{c}{ResNet34 (Excluding the last layer)} \\
encoder & \multicolumn{1}{l|}{Fixed Layer} & \multicolumn{2}{c}{Except for the ``Conv5'' block} \\ \midrule
Audio & \multicolumn{1}{l|}{Model} & \multicolumn{2}{c}{VGGish (Excluding the last layer)} \\
encoder & \multicolumn{1}{l|}{Fixed Layer} & \multicolumn{2}{c}{Except for the last CNN layer} \\ \midrule
\multirow{3}{*}{MultiTrans} & \multicolumn{1}{l|}{\# of Transformer block} & \multicolumn{2}{c}{2} \\
 & \multicolumn{1}{l|}{\# of head} & \multicolumn{2}{c}{4} \\
 & \multicolumn{1}{l|}{\# of hidden unit} & \multicolumn{2}{c}{256} \\ \midrule
\multirow{5}{*}{Guided-MELD} & \multicolumn{1}{l|}{$\kappa$ for video} & $0\leq\kappa\leq 2/M$ & $0\leq\kappa\leq 2/M$ \\
 & \multicolumn{1}{l|}{$\kappa$ for audio} & $0\leq\kappa\leq 2/N$ & $3/N \leq\kappa\leq 5/N$ \\
 & \multicolumn{1}{l|}{EMA decay} & $\rho=0.95$ & $\rho=0.999$ \\
 & \multicolumn{1}{l|}{$\lambda_0$} & 0.01 & 0.03 \\ 
 & \multicolumn{1}{l|}{$\gamma$} & \multicolumn{2}{c}{1.05} \\ \midrule
 & \multicolumn{1}{l|}{Optimizer} & \multicolumn{2}{c}{AdamW~\cite{adamw}} \\
Learning & \multicolumn{1}{l|}{Learning rate} & $0.001$ & $0.0005$ \\
parameter & \multicolumn{1}{l|}{L2 weight} & \multicolumn{2}{c}{0.001} \\
 & \multicolumn{1}{l|}{\# of maximum epoch} & \multicolumn{2}{c}{50} \\ \bottomrule
\end{tabular}
}
\end{table}

\begin{figure}[tb!]
 \begin{center}
 \includegraphics[width=0.88\linewidth]{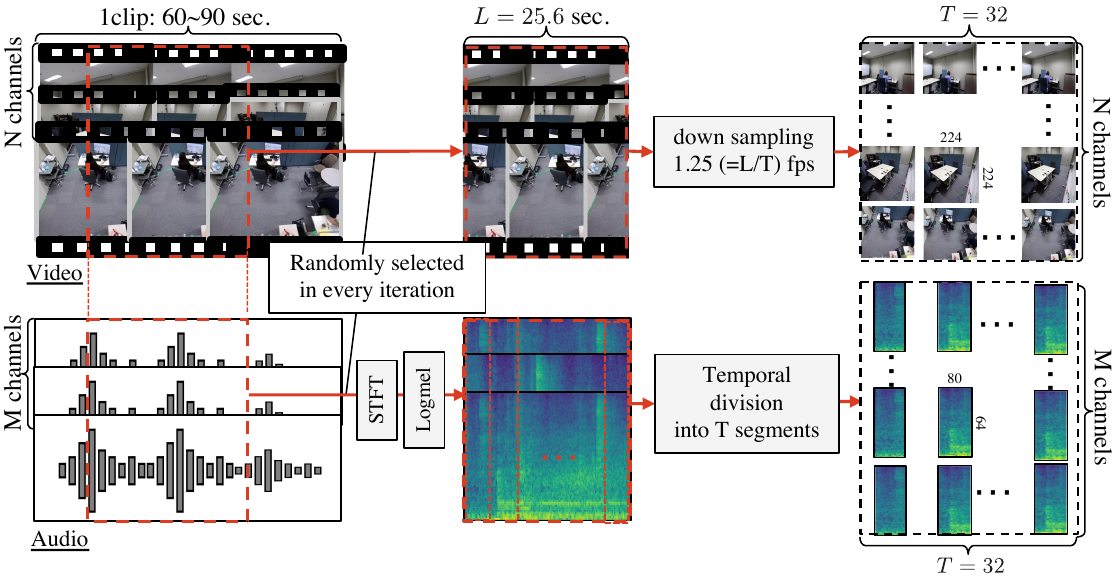}
 \caption{Shaping procedure of video and audio inputs during training. ``STFT'' and ``Logmel'' denotes the operation of a short-time Fourier transform of an acoustic signal and a extracting the absolute values of a log-mel spectrogram.}
 \label{fig:inputshaping}
 \end{center}
 \end{figure}

\subsection{Hyperparameters}
Table~\ref{tb:hyperparameter} lists the hyperparameters used in this experiment. These hyperparameters were determined by using the validation set. In the following, we describe the hyperparameters used in the experiment for the MM-Store dataset; for the MM-Office case, see this table.

Fig.~\ref{fig:inputshaping} shows the video and audio inputs shaping procedure. The input of the system is the audio signal obtained from the microphone with $M=12$ channels and the video signal obtained from the camera with $N=6$ channels. We fix the input clip length at $L=25.6$ seconds. Since each training data is about 30 to 90 seconds and longer than $L$, clips of length $L$ are randomly sampled from the training data at every iteration. The video input is downsampled from 30 to 1.25 ($=32.0/25.6$) fps and compressed to the resolution to 224 $\times$ 224. The sampling frequency of audio input was downsampled from 48 to 16 kHz, and the STFT spectrogram was extracted by using a 400-point Hanning window with a 160-point shift. The Mel-filter bank was 64 dimensions. The resulting Mel-spectrogram is then divided into 32 segments of 80 time frame lengths to align it with the frame rate of the video inputs.

The output dimensions of the audio and video encoder $\mathcal{A}_e$ and $ \mathcal{V}_e$ are both 512, and the embedding dimension is $D = 64$. We fixed all parameters except the final layer convolutional neural network (CNN) for the audio encoder $\mathcal{A}_e$. For the video encoder $\mathcal{V}_e$, we fixed parameters other than the Conv5 block. In MultiTrans, the Transformer block was stacked two layers, and the MHSA in the Transformer block had $H=4$ number of heads. In the preliminary experiments, the performance was degraded at a lower or higher number of heads $H$.

As shown Eq.~\ref{eq:mask2}, sensors are masked in accordance with the masking rate $\kappa$. In the experiment with the MM-Store dataset, each iteration was performed while varying $\kappa$ in the range $0\leq\kappa\leq 2/M$. In the experiment with the MM-Office dataset, a larger masking ratio range ($3/N \leq\kappa\leq 5/N$) was applied to the audio due to the high redundancy of the microphones. For the MM-Office dataset experiments, a smaller masking factor range ($0.03$) was applied to the video because of the lower camera redundancy. The EMA decay (Eq.~\ref{eq:ema}) in the parameter update of the Target network was set to $\rho=0.95$. The scheduling (Eq.~\ref{eq:schedule}) of the balance parameter $\lambda$ for multitask learning was performed with $\lambda_0=0.01$ as the initial value and $\gamma=1.05$ as the bottom of the exponent.

The AdamW optimizer~\cite{adamw} was used for all training, the initial learning rate was 0.001 and exponentially decayed to 0.1 times at the every epoch. The weight decay parameter was 0.001. The maximum number of epoch was 50. All training was concluded with the lowest loss value in the validation set.

\begin{table*}[htbp]
\centering
\caption{(a) Event tagging and (b) event detection scores on MM-Store dataset. The error range represents the standard error of each metric on three experiments with the different random seed. }
\label{tb:result_store}
\scalebox{0.95}[0.95]{
\begin{tabular}{l|cccc|cccc}
\toprule
 & \multicolumn{4}{c|}{(a) Event tagging score [\%]} & \multicolumn{4}{c}{(b) Event detection score [\%]} \\
 & \multicolumn{1}{c}{mAP} & \multicolumn{1}{c}{mAP (best)} & \multicolumn{1}{c}{ROAUC} & \multicolumn{1}{c|}{ROAUC (best)} & \multicolumn{1}{c}{mAP} & \multicolumn{1}{c}{mAP (best)} & \multicolumn{1}{c}{ROAUC} & \multicolumn{1}{c}{ROAUC (best)} \\ \midrule
(A) Late Fusion 
&$86.9\pm1.0$&$88.3$& $93.7\pm0.4$&$94.2$ 
&$32.1\pm0.2$&$32.5$&$88.5\pm0.1$&$88.7$ \\
(B) Concat 
&$81.7\pm0.5$&$82.8$& $92.8\pm0.9$&$94.1$
&$32.6\pm0.3$&$33.2$&$88.1\pm0.1$&$89.2$ \\
(C) CRF~\cite{crfpaper} 
&$84.5\pm1.9$&$87.0$& $93.8\pm0.5$&$94.7$
&$32.3\pm0.7$&$34.0$&$88.0\pm0.2$&$88.7$ \\
(D) MultiTrans~\cite{multitrans} 
&$84.6\pm1.1$&$87.2$& $94.7\pm0.2$&$95.0$ 
&$33.0\pm1.3$&$35.6$&$89.2\pm0.04$&$89.3$ \\
(E) MSM (M2D)~\cite{m2d} 
&$74.4\pm2.4$&$79.8$& $91.5\pm0.7$&$93.1$ 
&$28.5\pm1.4$&$31.8$&$86.1\pm0.6$&$87.6$ \\
(F) {\bf Ours: Guided-MELD} 
&${\bf 93.0}\pm1.4$&${\bf 96.5}$& ${\bf 97.1}\pm0.4$&${\bf 98.2}$
&${\bf 37.3}\pm1.2$&${\bf 39.7}$&${\bf 90.7}\pm0.3$&${\bf 91.2}$ \\
\midrule
(G) Guided-MELD w/o $\mathcal{L}_{M}$
&$83.8\pm2.7$&$89.6$& $93.6\pm0.8$&$95.2$
&$34.5\pm2.1$&$38.8$&$89.5\pm0.6$&$90.9$ \\
(H) Guided-MELD (CRF) 
&$64.3\pm2.1$&$69.3$& $85.9\pm0.6$&$87.0$ 
&$20.6\pm0.5$&$21.7$&$79.2\pm1.3$&$82.4$ \\ \bottomrule

\end{tabular}
} 
\end{table*}

\subsection{Results with MM-Store dataset}
\label{sec:resultstore}
This section describes the results of experiments using the MM-Store dataset.

\subsubsection{\bf Event tagging and event detection performance}
Table~\ref{tb:result_store} shows the event tagging and detection scores of the proposed method and the comparison methods. First in the (a) event tagging task, the proposed method (F) using Guided-MELD significantly outperformed the other methods (A) - (E) in all metrics. In the event tagging task, (B), (C), and (D) fell below the late fusion method (A). A possible reason for this is that this task is an easy problem even for the sensor-independent method (A) because identifying the class of an event only requires a brief capture of the target. Nevertheless, the fact that the proposed method (E) performed the best suggests that it captures events that can only be identified by linking multiple sensors well. 

Even in the more challenging task of (b) event detection, the proposed method (F) with Guided-MELD significantly outperformed the other methods (A)-(E) in terms of all metrics. In contrast to the event tagging task, late fusion (A) scored the worst, indicating that the inter-sensor relationship is more important to consider in the event detection task. It is considered that since identifying an event's onset/offset time requires constant knowledge of the target's behavior, it is difficult without cooperation among sensors. Multiple-sensor-based methods (B), (C), and (D) gradually improves the best scores of mAP and ROAUC compared with a method without sensor cooperation (A), but the average scores are comparable. The proposed method (F) outperforms (A) in terms of average and maximum scores, indicating that the inter-sensor relationship modeling with Guided-MELD effectively detects events in ever-changing situations.

Method (E) achieves moderately successful (a) event tagging and (b) event detection performance, despite not using event labels in the training of its encoder. This suggests that the extension to the distributed sensor of MSM is beneficial in extracting embeddings for event detection. However, it has lower performance than all other methods. This is likely because the distributed sensor observations contain numerous backgrounds, and M2D embeds them in the latent space without distinguishing between them.

The (a) event tagging and (b) detection scores of (G), which removes Eq.~\ref{eq:gmsm} from Guided-MELD, are worse than those of (F). This indicates that sensor-masking Eq.~\ref{eq:mask} alone does not improve performance due to data augmentation effects. This result leads to a different conclusion from the fact that random masking for images~\cite{maskaug1,maskaug2} or audio~\cite{SpecAugment} are also effective as data augmentation methods. From these results, in Guided-MELD, compensating for masked sensor information, rather than masking itself, plays an important role in modeling inter-sensor relationships. The (a) event tagging and (b) detection scores of (H) show a significant decrease in performance compared with (F) due to replacing the Multitrans with the CRF. This result indicates that a multi-sensor fusion mechanism for Guided-MELD needs a framework that can model changes in inter-sensor relationships depending on the situation, as MultiTrans does.

\begin{figure}[t!]
 \centering
 \includegraphics[width=0.88\linewidth]{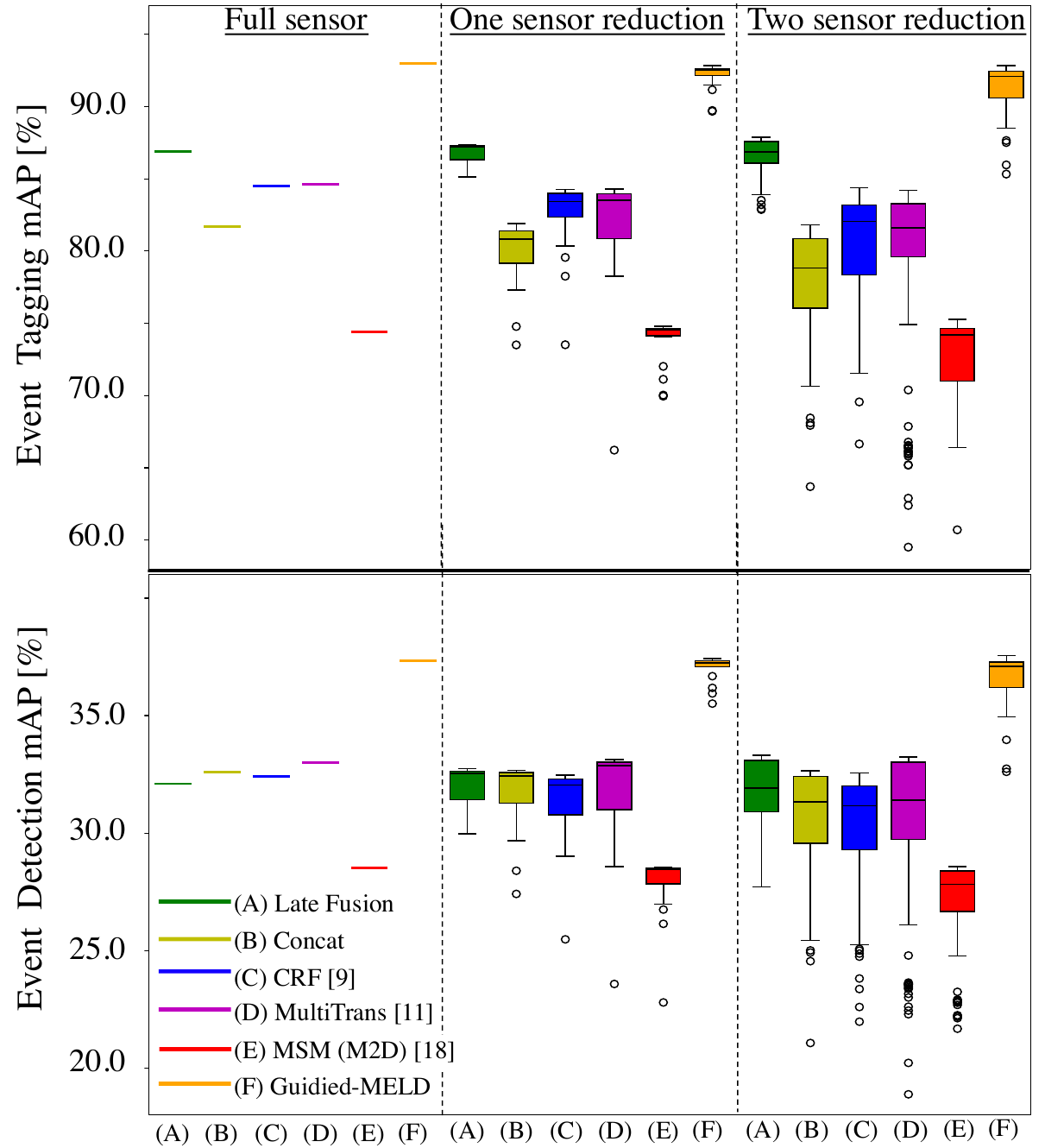}
 \caption{Sensor ablation study with MM-Store dataset. The box-and-whisker diagram shows the results of the experiment for all combinations that reduce one or two sensors. } 
 \label{fig:sensor_ablation}
\end{figure}

\begin{table}[t!]
\centering
\caption{Event detection performance on several guiding schedule strategies.}
\label{tb:schedule}
\scalebox{0.91}[0.91]{
\begin{tabular}{l|cccc}
\toprule
 & \multicolumn{2}{c}{Event tagging score [\%]} & \multicolumn{2}{c}{Event detection score [\%]}\\ 
 & mAP & ROAUC& mAP & ROAUC \\ \midrule
(F-1) Decrease $\lambda$ & $84.7\pm1.3$ & $94.0\pm0.6$ & $30.3\pm1.8$ & $87.4\pm0.4$ \\
(F-2) Constant $\lambda$ & $84.5\pm2.9$ & $93.2\pm1.4$ & $35.6\pm2.1$ & $88.3\pm0.7$ \\
(F-3) Increase $\lambda$ & $\bm{93.0}\pm1.4$ & $\bm{97.1}\pm0.4$ & $\bm{37.3}\pm1.2$ & $\bm{90.7}\pm0.3$ \\ \bottomrule
\end{tabular}
}
\end{table}

\begin{table*}[htbp]
\centering
\caption{(a) Event tagging and (b) event detection scores on MM-Office dataset. The error range represents the standard error of each metric on three experiments with different initial values.}
\label{tb:result_office}
\scalebox{0.95}[0.95]{
\begin{tabular}{l|cccc|cccc}
\toprule
 & \multicolumn{4}{c|}{(a) Event tagging score [\%]} & \multicolumn{4}{c}{(b) Event detection score [\%]} \\
 & \multicolumn{1}{c}{mAP} & \multicolumn{1}{c}{mAP (best)} & \multicolumn{1}{c}{ROAUC} & \multicolumn{1}{c|}{ROAUC (best)} & \multicolumn{1}{c}{mAP} & \multicolumn{1}{c}{mAP (best)} & \multicolumn{1}{c}{ROAUC} & \multicolumn{1}{c}{ROAUC (best)} \\ \midrule
(A) Late Fusion 
&$78.4\pm2.8$&$85.1$& $88.5\pm1.6$&$92.0$
&$40.8\pm1.8$&$44.2$&$85.8\pm0.9$&$87.9$ \\
(B) Concat 
&$81.0\pm1.2$&$83.7$& $90.8\pm0.4$&$91.7$
&$45.4\pm1.5$&$47.6$&$86.9\pm0.1$&$87.0$ \\
(C) CRF~\cite{danet} 
&$81.9\pm0.3$&$82.7$& $91.2\pm0.4$&$92.1$
&$46.5\pm0.2$&$47.0$&$87.0\pm0.3$&$87.6$ \\
(D) MultiTrans~\cite{multitrans} 
&$83.4\pm0.5$&$84.5$& $91.5\pm0.4$&$92.4$
&$46.8\pm0.3$&$47.3$&$86.5\pm0.3$&$86.9$ \\
(E) MSM (M2D)~\cite{m2d}
&$66.5\pm0.7$&$67.6$& $80.6\pm0.6$&$81.4$ 
&$31.6\pm0.5$&$32.3$&$77.8\pm0.7$&$79.4$ \\
(F) {\bf Ours: Guided-MELD} 
&${\bf 85.1}\pm0.8$&$86.7$& ${\bf 92.6}\pm0.2$&$93.0$ 
&${\bf 48.4}\pm0.6$&$49.7$&${\bf 88.0}\pm0.5$&${\bf 89.2}$ \\ \midrule
(G) Guided-MELD w/o $\mathcal{L}_M$
&${\bf 85.5}\pm0.8$&$87.2$& $92.3\pm0.3$&$92.9$
&$47.4\pm0.4$&$48.2$&${\bf 88.4}\pm0.3$&$88.9$ \\
(H) Guided-MELD (CRF) 
&${\bf 84.7}\pm1.2$&${\bf 87.5}$& ${\bf 92.4}\pm0.6$&${\bf 93.8}$
&${\bf 47.9}\pm1.3$&${\bf 50.9}$&$87.2\pm0.6$&$88.7$ \\ \bottomrule
\end{tabular}
}
\end{table*}

\subsubsection{\bf Sensor reduction experiment}
Figure~\ref{fig:sensor_ablation} is a boxplot showing the change in mAP score with sensor reduction. The proposed method (F) achieved a higher score than comparison methods in all sensor combinations. While method (A) demonstrates robustness to sensor reduction, its median performance consistently falls below that of the proposed method (F). Methods (B), (C), and (D) exhibit significant performance degradation with the reduction of certain sensor combinations (such as Cameras 0 and 1). This indicates these systems depend on a few sensors and fail to adequately capture event-related information upon their loss. Method (E) shows less performance degradation due to sensor reduction than methods (B), (C), and (D). This suggests the successful extraction of embeddings through MSM-based representation learning that is as invariant as possible to sensor reduction. However, as with the full-sensor case, median performance is consistently low. These results indicate that Guided-MELD is more effective at compensating for reduced sensor coverage through the collaboration of multiple sensors than the alternative methods.

In the proposed method (F), the maximum performance in each sensor number condition is almost constant. At the same time, in (A) to (D), there are combinations where the accuracy improves slightly with the reduction of sensors. One possible reason for this is that methods (A) to (D) do not fully utilize the information from all sensors, and event detection is hindered by redundant sensor information. In Guided-MELD, thanks to the joint optimization of Eq.~\ref{eq:gmsm} and Eq.~\ref{eq:gmsm2}), only the sensor information useful for event analysis can be extracted. Therefore, these results indicate that the proposed method eliminates the negative effects of redundant sensors.

This result offers promising real-world applications. For example, in the application of automatic detection of customer behavior in convenience stores that we focus on, it is desirable to use the same system in multiple stores. However, installing sensors at the same location in all stores is difficult because of the restrictions in each store. Even in this case, if the system is robust against sensor defects, it would be possible to train the system once with a sufficient amount of distributed sensors and then throttle back to the required amount for each store. Another point of view is the robustness to sensor failure. If the sensor fails, the cost of maintaining the system can be reduced if detection performance is maintained at a high level.

\subsubsection{\bf Investigation of guiding schedule}
Table~\ref{tb:schedule} shows the performance of event tagging and detection when varying the scheduling strategy for the loss function balance parameter $\lambda$ in multitask learning. The results show that (F-3), which increases $\lambda$ with learning as shown in Eq.~\ref{eq:schedule}, performs better on all metrics. On the other hand, score of (F-1) and (F-2) are comparable to that of the baseline method. This results suggest that by focusing on the downstream task in the early stages of learning and gradually guide Eq.~\ref{eq:gmsm} in the later stages, the guiding process of the proposed method works well.

\subsection{Results with MM-Office dataset}
To demonstrate the effectiveness of our method on a different dataset, we present the results of experiments conducted on the MM-Office dataset. Table~\ref{tb:result_office} shows the scores of (a) event tagging and (b) detection. Even in experiments using the MM-Office dataset, the proposed method (E) performed better than the other methods (A) - (D) in all metrics. The results show that the Guided-MELD works robustly even under given conditions with different numbers of room and sensor placement sensors.

One difference with the MM-Store dataset is that (G), which uses CRFs for the feature fusion mechanism in Guided-MELD, maintains (a) event tagging performance relative to (E). It indicates that the fixed parameters are sufficient for modeling the relationship between sensors in event tagging on the MM-Office dataset. One possible reason is that MM-Office handles smaller rooms and fewer sensors than MM-Store.

\section{Conclusion}
In this study, we newly approach distributed multimedia sensor event analysis (DiMSEA), an essential technique for capturing human or machine activities in complex and extensive environments such as a store. To successfully coordinate distributed multimedia sensors, we proposed Guided Masked sELf-Distraction modeling (Guided-MELD). Guided-MELD extracts a joint representation of distributed multimedia sensor observation that distills relevant information about the target event from observations of distributed sensors while ignoring obstructive information.

The newly recorded datasets MM-Store and MM-Office for the DiMSEA are the first large-scale datasets to record human activity by using distributed cameras and microphones. In numerical experiments on these datasets, the proposed method with Guided-MELD outperforms both the baseline and comparison methods. For the event tagging task, the mean average precision (mAP) score of the proposed method ($93.0\%$) is higher than that of the late fusion method, which is the best performing comparison method ($86.9\%$). In the event detection task, the mAP score of the proposed method ($37.3\%$) is higher than the score of conventional inter-sensor relationship modeling method conditional random fields (CRF; $32.3\%$) and the score of MultiTrans ($33.0\%$) proposed in our conference paper. These results show that Guided-MELD effectively linked distributed multimedia sensors to identify events in convenience store and office environments. Furthermore, the proposed method performed more robustly than the comparison methods when sensors were reduced. The robustness of the proposed method of sensor reduction offers high utility in practical applications, such as reduced sensor maintenance costs and high flexibility in sensor placement.

\bibliographystyle{IEEEtran}
\footnotesize
\bibliography{refs}

\appendices
\vfill

\end{document}